\documentclass[twocolumn, graphicx]{revtex4-1}

\usepackage{graphicx}
\usepackage{textcomp}
\usepackage{hyperref}
\usepackage{color}

\begin{document}

\title{Broadband, efficient extraction of quantum light by a photonic device comprised of a metallic nano-ring and a gold back reflector} 

\author{Cori Haws}
\affiliation{James Watt School of Engineering, Electronics \& Nanoscale Engineering Division, University of Glasgow, Glasgow G12 8LT, United Kingdom}
\affiliation{Physical Measurement Laboratory, National Institute of Standards and Technology, 100 Bureau Drive Gaithersburg, MD 20899, USA}

\author{Edgar Perez}
\affiliation{Physical Measurement Laboratory, National Institute of Standards and Technology, 100 Bureau Drive Gaithersburg, MD 20899, USA}
\affiliation{Joint Quantum Institute, NIST/University of Maryland, College Park, MD 20742, USA}

\author{Marcelo Davanco}
\affiliation{Physical Measurement Laboratory, National Institute of Standards and Technology, 100 Bureau Drive Gaithersburg, MD 20899, USA}

\author{Jin Dong Song}
\affiliation{Center for Opto-Electronic Materials and Devices Research, Korea Institute of Science and Technology, Seoul 136-791, South Korea}

\author{Kartik Srinivasan}
\affiliation{ Physical Measurement Laboratory, National Institute of Standards and Technology, 100 Bureau Drive Gaithersburg, MD 20899, USA}
\affiliation{Joint Quantum Institute, NIST/University of Maryland, College Park, MD 20742, USA}

\author{Luca Sapienza\\
E-mail address: \href{mailto:luca.sapienza@glasgow.ac.uk}{luca.sapienza@glasgow.ac.uk}\\
Website: \href{https://sites.google.com/view/integrated-quantum}{https://sites.google.com/view/integrated-quantum}}
\affiliation{ James Watt School of Engineering, Electronics \& Nanoscale Engineering Division, University of Glasgow, Glasgow G12 8LT, United Kingdom}
\affiliation{Physical Measurement Laboratory, National Institute of Standards and Technology, 100 Bureau Drive Gaithersburg, MD 20899, USA}


\begin{abstract}

To implement quantum light sources based on quantum emitters in applications, it is desirable
to improve the extraction efficiency of single photons. In particular controlling the
directionality and solid angle of the emission are key parameters, for instance, to
couple single photons into optical fibers and send the information encoded in quantum
light over long distances, for quantum communication applications. In addition, fundamental studies of the radiative behavior of quantum emitters, including studies of coherence and blinking, benefit from such improved photon collection. Quantum dots
grown via Stranski-Krastanov technique have shown to be good candidates for bright,
coherent, indistinguishable quantum light emission. However, one of the challenges
associated with these quantum light sources arises from the fact that the emission
wavelengths can vary from one emitter to the other. To this end, broadband light
extractors that do not rely on high-quality factor optical cavities would be desirable, so
that no tuning between the quantum dot emission wavelength and the resonator
used to increase the light extraction is needed. Here, we show that metallic nano-rings combined with gold back reflectors increase the collection efficiency
of single photons and we study the statistics of this effect when quantum dots are
spatially randomly distributed within the nano-rings. We show an average increase in the
brightness of about a factor 7.5, when comparing emitters within and outside the nano-rings in devices with a gold back reflector, we measure count rates exceeding 7\,$\times$\,$ 10^6$ photons per second and single photon purities as high as 85\,$\%$\,$\pm$\,1\,$\%$. These results are
important steps towards the realisation of scalable, broadband, easy to fabricate
sources of quantum light for quantum communication applications.

\end{abstract}


\maketitle 

Increasing the light extraction and controlling the optical mode the light is emitted into
and its angular distribution are key parameters for quantum light sources. Collection out
of the chip is particularly important if quantum light is used for quantum communication
by encoding information and transmitting it over long distance with optical fibers. The
implementation of broadband light extractors based on plasmonic resonators has been a focus
of research as a way to overcome the need for the mutual tuning of optical cavity resonances
and of the emission wavelength of a variety of light sources under study, including classical
and quantum emitters \cite{plasmonic_res}. If we focus our attention on semiconductor quantum dots, they
have shown to be sources of bright, indistinguishable and coherent quantum light \cite{QDs}. However
given that each emitter operates at slightly different wavelengths, each resonator needs to be
specifically designed to be on - or in most cases tuned into - resonance with the single emitter
under study \cite{Gammon, tuning}. To avoid this drawback, in particular if one considers using such emitters in applications, scalable broadband light extractors are desirable, as is bright photon-pair emission \cite{JinXX}. To this end, we have
shown that metallic nano-rings allow increasing the extraction efficiency of single photons up
to a factor 20 \cite{rings} and that they can be combined with super-solid immersion lenses to reach
photon fluxes up to 1\,$\times$\,$ 10^6$ photons per second \cite{SIL}. The collimating effect provided by the metallic nano-ring also helps reducing the effective numerical aperture (NA) of the emission cone, thus improving the coupling to low-NA single-mode optical fibers.

\begin{figure}[]
\centering
\includegraphics[width=1\linewidth]{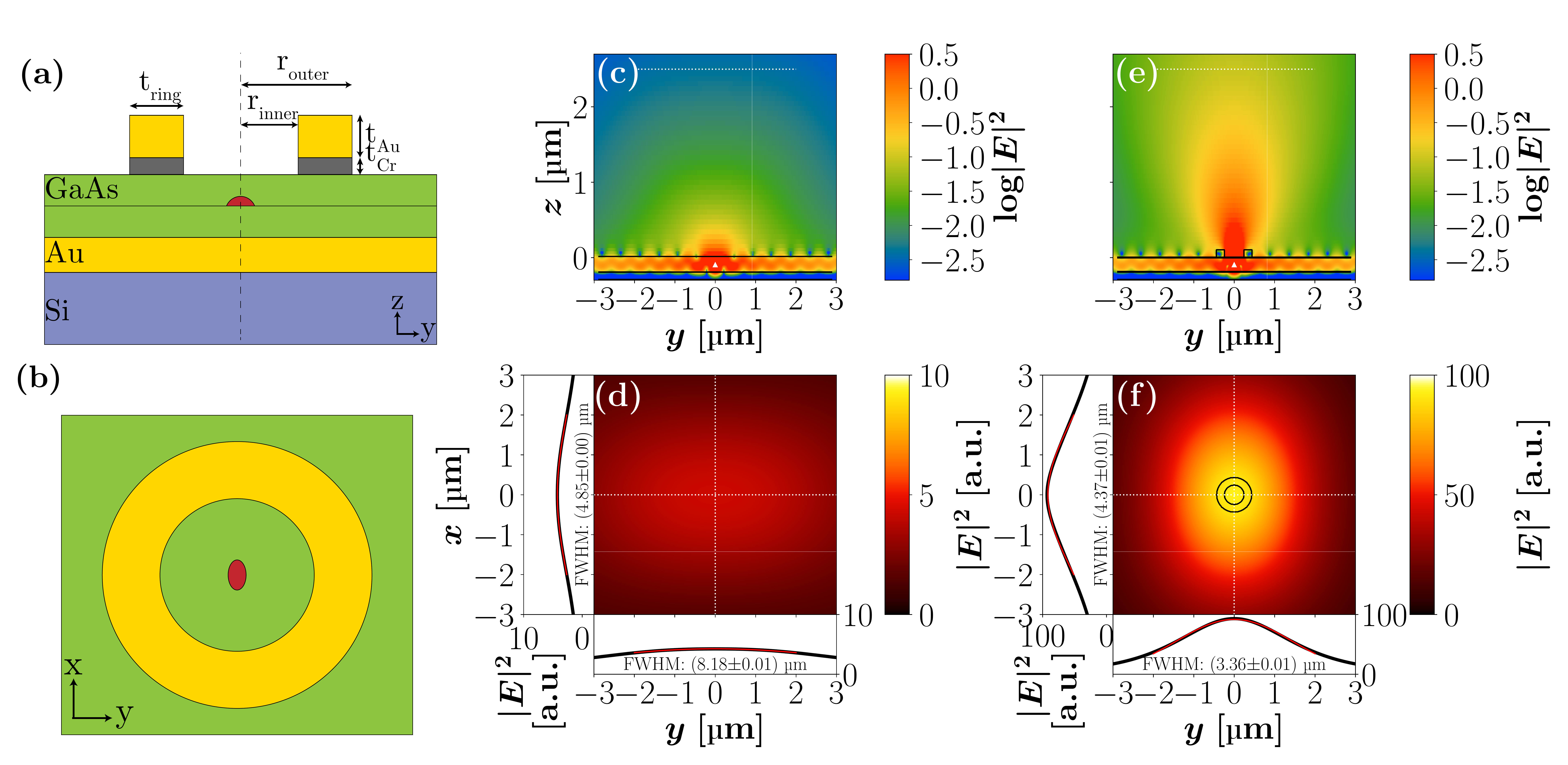}
\caption{(a) Schematic (not to scale) of the metallic nano-ring with Au back reflector device geometry. The Au back reflector, deposited onto the silicon substrate by electron-beam assisted metal
evaporation, is 100\,nm thick, the GaAs membrane is 190\,nm thick and has a layer of quantum dots
in the middle (a single quantum dot is depicted by the red shape). The optimised ring has the following dimensions: r$_{inner}$\,=\,236\,nm, r$_{outer}$\,=\,437\,nm, t$_{ring}$\,=\,201\,nm, t$_{Cr}$\,=\,7\,nm, t$_{Au}$\,=\,93\,nm. (b) Top view of the nano-ring device shown in panel (a). (c) Colour plot of the side ($zy$ plane) profile of
the squared electric field of a dipole emitter (oriented along the $x$ axis and emitting between 925 and 975\,nm), obtained by finite-difference time-domain simulations, and (d) top view ($xy$ plane), measured at a distance of 2.5\,$\mu$m from the sample surface in correspondence to the dotted line shown in panel (c), including linecuts of the far-field profile (black lines) and their Gaussian fits (red lines). The Full-Width Half-Maximum (FWHM) obtained from the
fits is indicated in the graphs, with the error obtained from the fitting function (one standard deviation uncertainty due to the fit negligible compared to the values). (e, f) Same as panels (c, d) in the presence of a metallic nano-ring (with dimensions listed above) on the sample surface, centered around the emitting dipole. The concentric black circles in (f) represent the nano-ring location on the sample surface.}
\end{figure}

Here, we show that metallic back reflectors, combined with nano-rings, can be used to collect some of the light emitted towards the substrate that otherwise would be lost when collection is carried out above the chip. The gold layer reflects part of the light emitted by quantum dots embedded within a GaAs membrane towards the collection objective and this metallic nano-ring-back reflector device geometry allows us to measure single-photon fluxes exceeding 7\,$\times$\,$ 10^6$ photons per second.
The sample consists of a high-density Stranski-Krastanov InAs/GaAs quantum dot layer, grown by molecular beam epitaxy, placed in the middle of a 190\,nm\,-\,thick GaAs membrane that is bonded to a Au layer on a silicon substrate \cite{membrane}, as schematically shown in Fig.\,1\,(a,b).
The dimensions of the metallic nano-rings, deposited onto the surface of the sample, are optimised by finite-difference time-domain simulations to maximise the amount of light emitted in the upwards direction towards the collection optics, as shown in Fig.\,1\,(c-f). Compared to the bulk structure (that has the Au back reflector but no nano-ring on the surface), an increase of about a factor 14 in the vertically extracted power (integrated over a 2\,$\mu$m\,$\times$\,2\,$\mu$m area) is observed, as well as
a collimating effect (obtained by considering the change in the full-width half-maximum of the
electromagnetic field) exceeding 40\,$\%$, thanks to the presence of the metallic nano-ring. As
discussed in previous papers \cite{rings,SIL}, these effects are broadband and work for emission wavelengths covering tens of nanometers, encompassing the entirety of the inhomogeneous distribution of the quantum dots in this sample. It is important to underline that a broadband enhancement of single-photon collection is important, for instance, to be able to investigate fundamental properties of quantum dots as-grown (like their coherence, recombination mechanisms, charge confinement and transport), without requiring etching of the surrounding material to increase the collected photon flux, a process that could impact the quantum dot properties \cite{etch, blink} or tuning of their emission wavelength. Other broadband approaches have relied on the deposition or deterministic fabrication of solid-immersion lenses \cite{SILs, det_SIL, QD_SIL} that require more complex fabrication processes, nano-antennas and plasmonic structures \cite{antenna, plas} that can provide less directionality of the emission.

\begin{figure}[]
\centering
\includegraphics[width=1\linewidth]{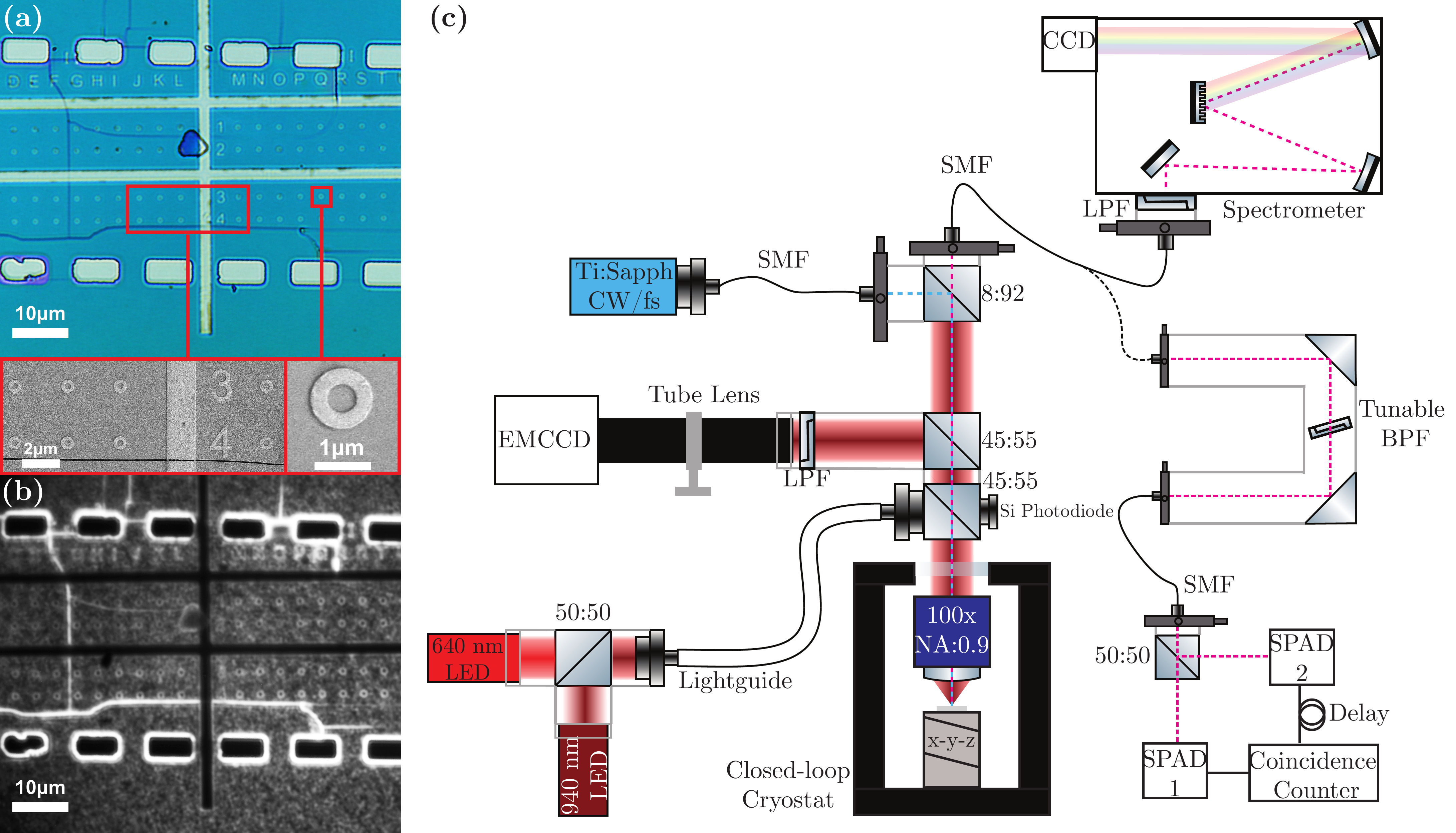}
\caption{(a) White-light image of a GaAs membrane (blue area) deposited on a Au layer (pale green
areas), with metallic markers and arrays of nano-rings on its surface. The bottom panels show scanning electron micrograph images of the zoomed-in areas delimited by the red rectangles. (b) Photoluminescence image of the same membrane shown in panel (a), collected under 640\,nm-light emitting diode (LED) illumination at a temperature of about 4\,K, showing the emission of InAs/GaAs quantum dots. (c) Schematic
of the confocal micro-photoluminescence setup (see main text for more details). LPF: long-pass
filter, BPF: band-pass filter, SMF: single-mode fiber, SPAD: single-photon avalanche detector, each square with a diagonal line represents a beamsplitter whose splitting ratio is shown as reflection:transmission next to it.}
\end{figure}

Figure 2\,(a) shows an optical image of one of the samples under study: the
blue area corresponds to the GaAs membrane embedding quantum dots; the green areas
are the gold layer, the markers and the array of metallic nano-rings, obtained via electron-beam
lithography, followed by metal lift off process (more details of the fabrication process are provided in Ref. \cite{rings}). Figure 2\,(b) shows a photoluminescence image of the quantum dot emission \cite{NComm, Jin}
that allows an estimation of the density of quantum dots: as shown, the metallic nano-rings contain more than one emitter and the location of the emitters with respect to the
center of the ring is uncontrolled. Figure 2\,(c) shows a schematic of the confocal micro-photoluminescence setup in use: the sample can be illuminated with two light-emitting
diodes (LEDs) to obtain the images shown in panels (a,b), by using an electron multiplying
charge-coupled device (EMCCD), and can be excited by a tunable Ti:Sapph laser operating
in continuous wave or with femto-second light pulses for time-resolved measurements. The signal, collected by an objective with NA\,=\,0.9, is spectrally characterised with a 0.5\,m long spectrometer equipped with a CCD or, after spectral filtering a selected emission line, sent to a Hanbury-Brown and
Twiss interferometer for single-photon purity characterisation or decay dynamics measurement, when one of the arms is blocked.

\begin{figure}[]
\centering
\includegraphics[width=1\linewidth]{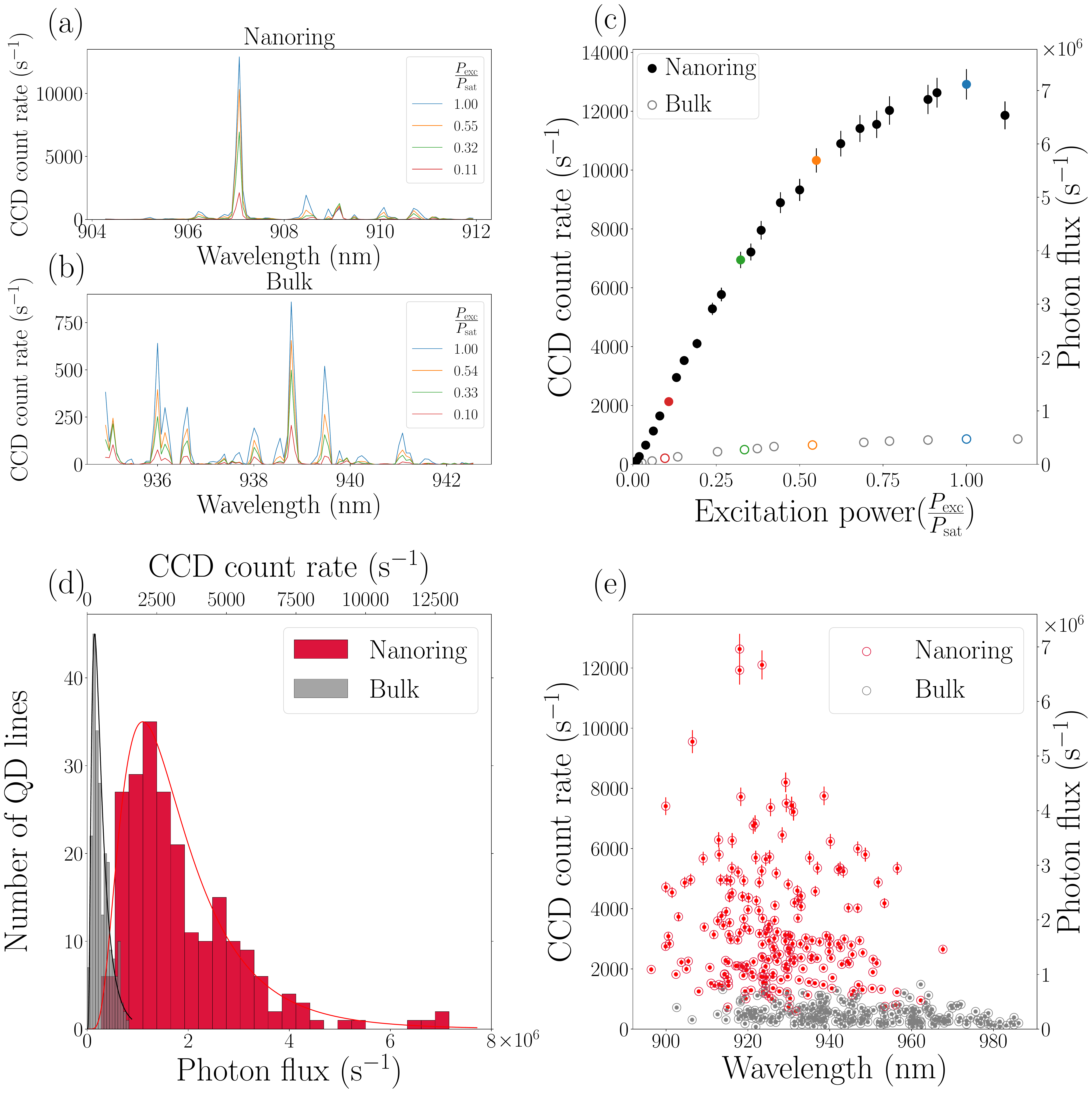}
\caption{(a,b) Example of photoluminescence spectra, collected at a temperature of about 4\,K, on
the CCD placed at the exit of the spectrometer, of quantum dots placed within (a) and outside (b)
nano-rings respectively. (c) Photoluminescence intensity of the emission lines with wavelength of
907.06\,nm in panel (a) and 938.78\,nm in panel (b) (colour coded accordingly), as a function of excitation power of a continuous wave 780\,nm laser (P$_{exc}$), normalised to the power giving the highest emission count rate for a specific emission line (P$_{sat}$\,=\,260\,$\mu$W). The photon flux corresponds to the number of single photons collected by the objective, once taking into consideration the losses in the collection path and the sensitivity of the detectors (correction factor of 550\,$\pm$\,22). Error bars represent one standard deviation values due to uncertainty in the loss calibration, and are shown when larger than the data point size. (d) Statistics of the collected emission intensity for quantum dots in bulk (outside the nano-rings) and within nano-rings (gray and red histograms
respectively), each composed of about 230 emission lines. The solid lines are log-normal fits to the
histograms. (e) Photoluminescence intensity of the emission lines from quantum dots within (red symbols) and outside (gray symbols) metallic nano-rings, plotted as a function of emission wavelength.}
\end{figure}

We characterise the emission of quantum dots in the unpatterned area of the chip and
within metallic nano-rings. Figure 3\,(a,b) shows examples of the emission spectra of the quantum
dots excited by a 780\,nm wavelength laser spot of about 1\,$\mu$m in diameter, at a temperature
of about 4\,K, and panel (c) shows the intensity of selected emission lines collected when
varying the continuous-wave laser excitation power, displayed in saturation power units:
a clear increase in the brightness of more than a factor 20 is observed. Given that the
emitters are randomly distributed within the nano-rings, we study the statistic of the collected
emission intensity for about 450 quantum dot emission lines, about half from quantum dots located outside the rings and half within the rings. The results are shown in panel (d): the log-normal distributions peak
at about 270 photons per second and at about 2000 photons per second outside and within the rings respectively and we measure single-photon fluxes exceeding 7\,$\times$\,$ 10^6$ photons per second, immediately after the collection objective (calculated by accounting for the losses in the collection path and the detector's sensitivity). These results illustrate the effectiveness of the metallic nano-ring and back reflector device geometry that we have developed in the improvement of the extraction of light emitted by solid-state sources.
It is worth noting that no wavelength tuning of the emitters was needed to obtain the increased brightness that we have reported, thus making the process scalable.\\
We take advantage of the fact that many single quantum dots are present within the nano-rings to experimentally verify the broadband nature of the effect under study. Panel (e) in Fig.\,3 shows the count rates measured on different quantum dot lines for emitters lying within and outside the nano-rings: as shown, the increased brightness, for a given nano-ring design, occurs over tens of nanometers, proving the broadband effect of the combined nanoring-back reflector device.

\begin{figure}[]
\centering
\includegraphics[width=1\linewidth]{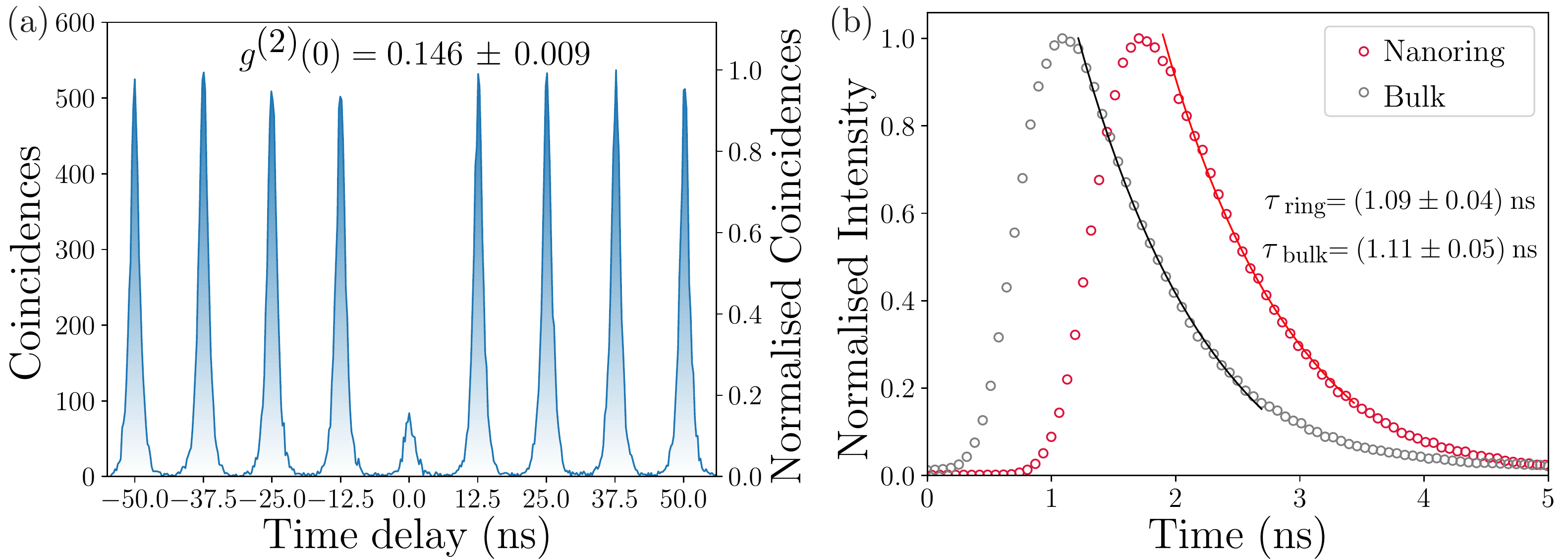}
\caption{(a) Single-photon correlation measurements carried out with a Hanbury-Brown and Twiss
setup, under femto-second pulsed, quasi-resonant excitation, of a spectrally filtered emission line
from a selected quantum dot placed within a metallic nano-ring. The value of the correlation
function $g^{(2)}(0)$, extracted by measuring the area of the peak at delay zero compared to the average
area of the surrounding peaks is shown. The error is calculated from the propagation of the error
in the evaluation of the areas and the standard deviation of the mean of the area of the peaks
at non-zero time delay. (b) Spontaneous decay dynamics of spectrally-filtered quantum dot lines
within and outside the nano-rings (red and gray symbols respectively) and their exponential
fits (solid lines). The time characteristics are displayed (lifetimes $\tau$) and
the errors are calculated from the fitting accuracy. One standard deviation uncertainties on the data points (due to fluctuations in the accumulated counts) are negligible, that is, smaller than the data point size.}
\end{figure}

We then characterise the single-photon purity and decay dynamics of the quantum dot
lines, as shown in Fig.\,4. We excite the quantum dots within a metallic nano-ring using a
filtered femtosecond laser tuned to a quasi-resonant excitation wavelength (884.7\,nm for a quantum dot line at 909.0\,nm): as shown in panel (a), a single-photon purity of about 85\,$\%$ is measured, proving that, despite the high density of emitters, the device under study allows good quality single-photon collection, after spectral filtering of single emission lines. As expected for quantum dots relatively
far from the metallic nano-ring and back reflector (95\,nm), the coupling to the plasmonic field
is not enough to modify the spontaneous emission dyanamics and, panel (b) shows that the
lifetimes for representative quantum dots outside and within the ring are both, as expected
for InAs/GaAs quantum dots, about 1\,ns, and the same within the one standard deviation uncertainties resulting from the nonlinear least squares fits to the data. Plasmonic devices
have shown to allow very strong reductions of spontaneous emission lifetimes \cite{lifetimes} but require
emitters closer (down to a few nanometers) to the metallic features (work in this direction is in progress \cite{droplet}).

In conclusion, we have shown that metallic nano-rings combined with gold back reflectors
increase the brightness of single-photons emitted by quantum dots. Our results
go towards the realisation of scalable, broadband photonic devices that do not rely on spatial positioning of emitters and/or on mutual tuning of emission wavelength and cavity resonances. Despite providing lower performances compared to state-of-the-art devices like micro-pillars \cite{pillars} and circular Bragg gratings \cite{Marcelo}, they have the advantage of having more broadband characteristics. Besides quantum photonic applications, the broadband brightness increase also makes this device geometry useful for the characterisation of fundamental properties of quantum and classical emitters, without requiring (or before they are embedded within) etched nanostructures that can deteriorate the emitter coherence properties and/or emission stability, due to the proximity of defects in the etched surfaces. 
Furthermore, these devices are produced via a simple fabrication process, not requiring any etch step, and therefore provide easy on-chip device integration, useful for photonic quantum information applications.

\section*{Acknowledgments}

LS acknowledges financial support by the Leverhulme Trust, grant IAF-2019-013.
JDS acknowledges the support from the KIST institutional Program and IITP grant funded by MSIT (No.20190004340011001).

\end{document}